# Electronic-Photonic Interface for Multiuser Optical Wireless Communication

Youngin Kim, Laurenz Kulmer, Jae-Yong Kim, Hamza Kurt, Juerg Leuthold, *Fellow, IEEE*, and Hua Wang, *Fellow, IEEE*

*Abstract*—We demonstrate an electronic-photonic (EP) interface for multiuser optical wireless communication (OWC), consisting of a multibeam optical phased array (MBOPA) along with co-integrated electro-optic (EO) modulators and high-speed CMOS drivers. The MBOPA leverages a path-length difference in the optical phased array (OPA) along with wavelength-division multiplexing technology for spatial carrier aggregation and multiplexing. To generate two and four pulsed amplitude modulation signals, and transmit them to multiple users, we employ an optical digital-to-analog converter technique by using two traveling-wave electrode Mach-Zehnder modulators, which are monolithically integrated with high-speed, wide-output-swing CMOS drivers. The MBOPA and monolithic EO modulator are implemented by silica wafer through planar lightwave circuit fabrication process and a 45-nm monolithic silicon photonics technology, respectively. We measured and analyzed two-channel parallel communication at a data rate of 54 Gbps per user over the wireless distance of 1 m. To the best of our knowledge, this is the first system level demonstration of the multi-user OWC using the in-house-designed photonic and monolithically integrated chips. Finally, we suggest best modulation format for different data rate and the number of multibeams, considering effects of the proposed OPA and the monolithic modulator.

## I. INTRODUCTION

Traditional radio frequency (RF) and mm-wave technologies are facing challenges for future wireless communication systems that require high data rates and low latency, mostly due to the congested microwave frequency spectrum [1]. In particular, the ongoing trend towards data-demanding applications, such as the Internet of Things (IoT) and artificial intelligence (AI), are driving the need for beyond 5G and sixth-generation (6G) wireless systems to achieve Gigabit-per-second (Gbps) or even Terabit-per-second (Tbps) data rates, along with intense data exchange between users, thereby imposing a greater burden on the limited spectrum resources.

To address this issue, multibeam antennas (MBAs) have attracted a significant interest in the mm-wave and THz spectrum [2, 3]. Thanks to their capability of producing a large number of concurrent beams, MBAs provide dynamic networking and massive multiple input multiple output (MIMO), allowing current RF wireless systems to overcome the scarce radio spectrum. Various methods, including passive MBAs (PMBAs) [4, 5], MB phased array antennas (MBPAAs) [6, 7], and digital MBAs (DMBAs) [8, 9], have been proposed to create multiple RF beams. Even though the RF MBAs mitigate the congested spectrum problem, they are typically very bulky, unscalable, or the performance is limited by the computing power of digital signal processing (DSP) chips [10, 11]. In contrast, optical wireless communication (OWC) systems provide an almost unlimited spectral range and enable high data rate and low latency communications for wireless links as well as a compact footprint, thanks to the higher carrier frequency of the light compared to RF and mm-wave frequencies [12–16].

To implement the OWC link, it is essential to develop a device that can provide beam forming and steering functions. Conventional OWC systems use spatial light modulator (SLM) [17–19] and micro-electromechanical system (MEMS) [20, 21] to create and control a collimated beam. However, SLM and MEMS are bulky, power hungry, and have limited steering angle, resulting in a blind area in wireless links and inapplicability to MB communications systems. Fiber arrays combined with a lens [22] or metasurfaces [23, 24] have also been proposed for the beam collimation and control. Despite their strong light manipulation and MB generation capabilities, their discrete beam-steering angle creates blind spots in OWC systems. Although the pair of arrayed fiber and metasurface in combined with wavelength division multiplexing (WDM) technique [24] demonstrated full-range coverage, it sacrifices beam divergence angle and requires additional polarization controllers for each fiber.

Manuscript received MM DD, YYYY; revised MM DD; accepted MM DD, YYYY. Date of publication MM DD, YYYY; date of current version MM DD, YYYY. This work was supported by 45SPCLO University Program from GlobalFoundries (Corresponding author: Youngin Kim).

Youngin Kim, Laurenz Kulmer, Juerg Leuthold and Hua Wang are with the Department of Information Technology and Electrical Engineering, ETH Zürich, 8092 Zürich, Switzerland (e-mail: youngkim@iis.ee.ethz.ch; hua.wang@iis.ee.ethz.ch)

Jae-Yong Kim and Hamza Kurt are with the School of Electrical Engineering, Korea Advanced Institute of Science and Technology (KAIST), 291 Daehak-Ro, Yuseong-Gu, Daejeon 34141, Republic of Korea.

A promising alternative is integrated optical phased arrays (OPAs) that perform beam forming and steering through on-chip arrayed optical antennas and phase shifters [25–39]. In contrast to conventional beam manipulators, the OPAs not only have continuous beam-steering angles but also provide extremely narrow divergence angle [25, 26], wide field-of-view (FoV) [27–30], compact size and improved energy and cost efficiencies [25–39] making them well-suited for in-door wireless communication links. There have been considerable efforts to realize high performance OPAs for OWC applications [31–36]. Four-element OPA fabricated on the silicon photonics platform showed 12.5 Gbps data transmission over a 1.4 m distance [31]. The longest chip-to-chip OWC link (50 m) has been demonstrated by a 512-element OPA with high-directionality waveguide grating antennas [32]. Recently, [35] achieved 50 Gbps four-level pulse amplitude modulation (PAM4) signal transmission in free-space by using a 128-element OPA, and [36] implemented the first OWC transmitter (TX), which monolithically integrates 64-element OPA, traveling-wave electrode Mach-Zehnder modulator (TWE-MZM), and high-speed CMOS driver all in one-chip.

However, conventional integrated OPAs designed for infrared light have not yet leveraged the enormous advantages of spatial multiplexing through the generation of MB for OWC applications. To achieve the full potential of the integrated OPA and to further boost the advantages of the OWC, which support high data rate, low latency, spectral efficiencies, it is necessary to communicate with independent dynamic beams which will enable massive MIMO [1–3, 24]. Various integrated OPAs have been proposed to form MB in free-space by employing Butler matrix [37], phase control [38], and polarization multiplexing [39]. However, the Butler matrix is unscalable as the number of 3 dB couplers and phase shifters significantly increases along with the required number of beams. Moreover, controlling phase shifters in large-scale OPAs require memory intensive look-up tables. Recently, a 64-element silicon nitride OPA demonstrated MB OWC with 1 Gbps data rate per each channel by multiplexing blue and green lights and rotating the polarizations [39]. However, visible light is incompatible with electro-optic (EO) modulators and photodetectors fabricated in typical advanced silicon photonics or III/V platforms that can support high data rate.

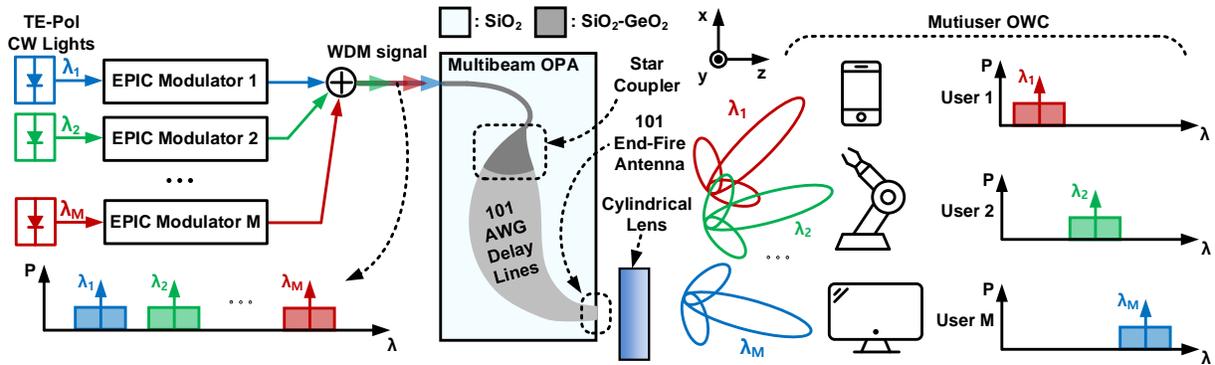

**Fig. 1.** An architecture of the multiuser optical wireless communication (OWC) using the proposed electronic photonic integrated circuit (EPIC) modulators and multibeam optical phased array (MBOPA).

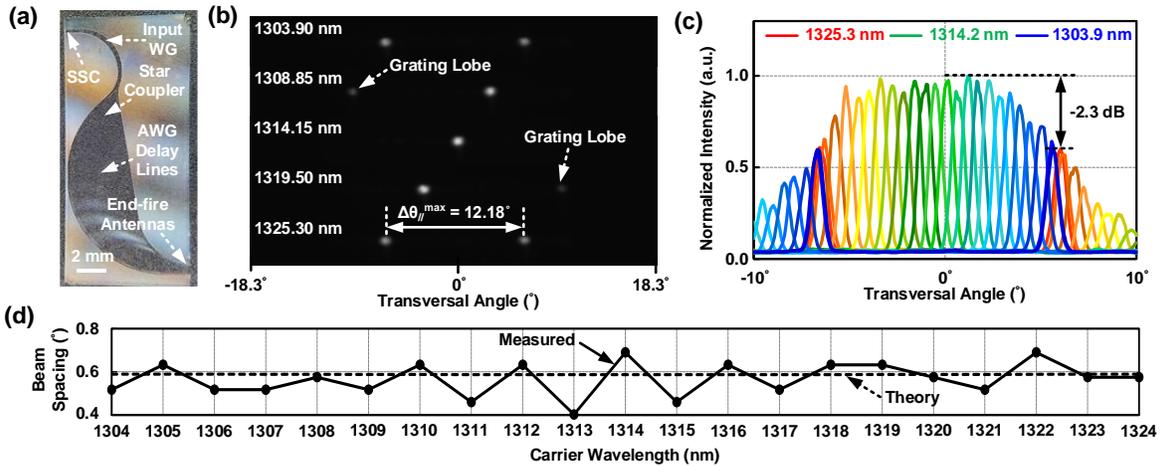

**Fig. 2.** Proposed silica-based multibeam optical phased array (MBOPA) with arrayed waveguide grating (AWG) delay lines. (a) Chip photograph of the proposed MBOPA. Measurement results of the proposed MBOPA: (b) Far-field patterns; (c) Normalized intensity of the MBOPA over the various carrier wavelengths; (d) Beam spacings plotted for the different carrier wavelengths.

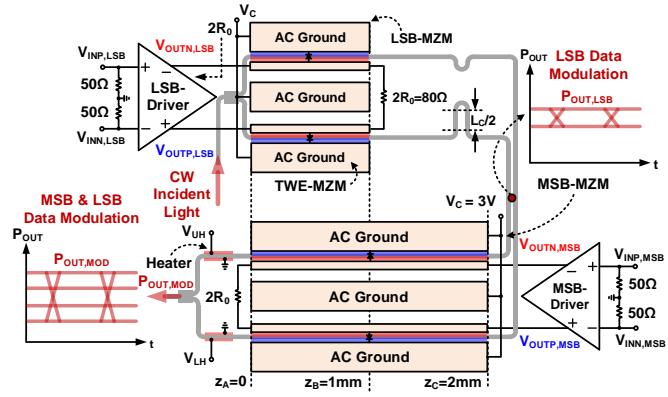

**Fig. 3.** A block diagram of the proposed co-integrated 2-bit optical digital-to-analog converter (ODAC)-based traveling-wave electrode Mach-Zehnder modulator (TWE-MZM) and high-speed CMOS cascode push-pull driver.

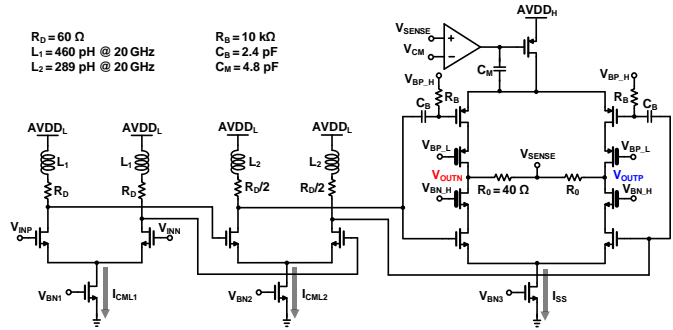

**Fig. 4.** Circuit schematics of the designed most significant bit (MSB) and least significant bit (LSB) drivers.

In this article, we propose a new electronic-photonic (EP) architecture that enables simultaneous OWC with multiple users. In the proposed architecture, a silica-based MBOPA with arrayed waveguide grating (AWG) delay lines combined with the WDM technique is introduced to generate MBs in free-space. Through simple AWG delay lines in the proposed OPA, the multiplexed high-speed signals in each carrier wavelengths are mapped into multiple narrow beams in free-space, allowing high-density dynamic networking. Thanks to the wavelengths in the micrometer range, the size of the MBOPA and divergence angle of the generated beams are dramatically smaller than MB antennas in the RF/mm-wave frequencies. Beam steering and wireless communications using the silica OPA by tuning C-band lasers has been introduced before [40]. However, application of the WDM technique for the simultaneous multi-user communication is presented for the first time in this proposed EP architecture. Moreover, we demonstrate new high-performance EO modulator realized in an electronic-photonic integrated circuit (EPIC) technology. Using the EPIC technology, TWE-MZMs were employed to realize an optical digital-to-analog converter (ODAC) scheme co-integrated with mm-wave CMOS push-pull drivers. Our proposed EPIC modulator can be programmed to generate pulsed amplitude modulation two or four (PAM2/4) signals. By using the proposed EPIC modulator, the PAM2/4 communication performances in free-space channels created by MBs have been measured and analyzed. The proposed MB generation technique using AWG delay lines with WDM is compact and highly scalable. Moreover, our MBOPA and EPIC modulator achieve highest performance, compared to state-of-the-art OPAs and EO modulator circuits.

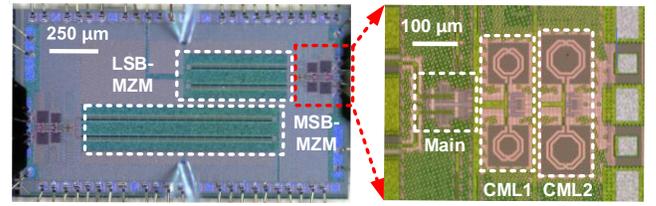

**Fig. 5.** Chip photograph of the proposed electronic-photonic integrated circuit (EPIC) modulator.

Section II presents the overall architecture of the proposed EP interface for OWC, along with the detailed implementation and measurement results of the in-house-designed MBOPA and EPIC modulator. Section III discusses the experimental results and Section IV concludes the paper.

## II. System Implementation and Results

### A. Proposed Architecture for multiuser OWC

Figure 1 shows an overall architecture of the proposed EP interface for multi-user OWC. Transverse-electric (TE) polarized continuous wave (CW) lights with different wavelengths are fed into the proposed M-element high-performance EPIC modulators. Then, the multiplexed modulated signals are coupled via a spot-size converter in our MBOPA. The star coupler splits the multiplexed signal in the 101-elements of the AWG delay lines with a path length difference of $\Delta L$. The modulated signals in the multiple carrier wavelengths are directed to multiple beams in free space through the wavelength-dispersive OPA principle, where the $\Delta L$ between neighboring channels allows beam direction to be controlled by wavelength variation, thereby enabling the multiple beam emissions in MBOPA. To reshape the diverging emitting beams forming an end-fire antenna over the longitudinal direction, a cylindrical lens with a focal distance of 10 mm is employed. These MBs that carry different data can be wirelessly transmitted to M-users, enabling massive MIMO communication in free-space optics.

The chip photograph of the proposed MBOPA is shown in Fig 2(a). The MBOPA is manufactured on a silica planar lightwave circuit platform with a 2.0% low-index contrast. The device has a total footprint of 20 mm × 9 mm, which includes 101-elements of end-fire antennas, and an estimated insertion loss of 2 dB, which are tremendously smaller than conventional RF and mm-wave MB antennas [2–9]. Benefiting from the platform's low-loss characteristics and wide transparency window, the fabricated MBOPA ensures high compatibility with EPICs operating in the O-band, while maintaining its previously validated standalone performances in the C-band [41]. Given a path difference of $\Delta L = 53.1 \ \mu m$ and a radiator pitch of d = 6 $\mu m$, the maximum angular range for enabling multi-user access is calculated to be 12.6° and it can be attainable through a wavelength variation of 21.4 nm, theoretically.

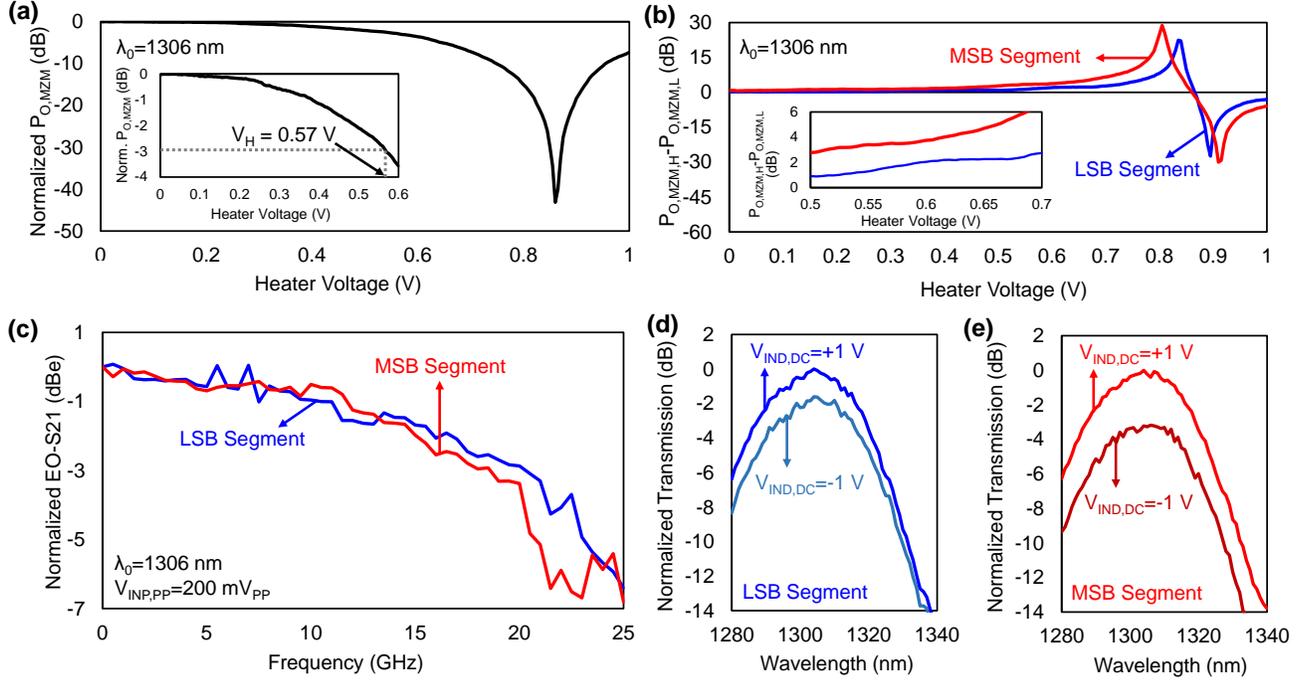

**Fig. 6.** DC and AC measurement results: (a) Normalized output power of the proposed electronic-photonic integrated circuit (EPIC) modulator over the varying heater voltage, when $V_{INP,MSB} = V_{INN,MSB} = V_{INP,LSB} = V_{INN,LSB} = 0.9$ V; (b) Optical power difference ($P_{OMZM,H}$-$P_{OMZM,L}$) between the high and low level of the output of the EPIC modulator over the heater voltage, at $\lambda_0 = 1306$ nm; (c) Normalized EO-S21 of the each modulator segments; Normalized transmission of the EPIC modulator when $V_{IND,DC} = +1$ and -1V are applied to the (d) most significant bit (MSB) and (e) least significant bit (LSB) driver.

Figures 2(b)-(c) present the measured far-field patterns in free-space and the characterized beam intensity distributions, obtained by incorporating a cylindrical lens in front of the edge-emitting MBOPA while varying the wavelength from 1303.90 nm to 1325.30 nm. The results show that beam transmission to multiple-users within a 12.18° range is available, closely aligning with theoretical expectations (for detailed theory, see [41]). During the wavelength tuning, grating lobes from the uniform-pitched radiators emerged, accompanied by an inherent OPA-induced degradation of main beam intensity as the phase difference shifted from 0 to ±π. Exhibiting a similar trend as reported in [41]. The measured beam angular shift per 1 nm wavelength variation is shown in Fig. 1(e), which, despite of slight measurement discrepancies, remains in close agreement with the theoretical value of 0.59°. The beam divergence angles over the transversal and longitudinal angles, $\theta_{//}$ and $\theta_{\perp}$, are 0.793° and 0.18°, respectively.

### B. Proposed EPIC Modulator

Figure 3 shows a block diagram of the proposed co-integrated ODAC-based TWE-MZM and the CMOS cascode push-pull drivers, which are designed and fabricated in a 45-nm monolithic silicon photonics technology by GlobalFoundries (45SPCLO) [42]. The proposed EPIC modulator consists of two segments for its modulation: A least significant bit (LSB) segment and a most significant bit (MSB) segment. The LSB and MSB segments contain the 1 mm and 2 mm-long $p$-$n$ junction phase shifters, respectively, which are driven by the co-integrated CMOS drivers. Since the silicon photonic devices in 45SPCLO technology are optimized for O-band, a TE-polarized CW light in this wavelength range is coupled into an input grating coupler (GC) of the EPIC modulator. Then, two electrical PAM2 signals, each containing LSB and MSB data, are fed into the LSB and MSB segments. The MSB segment modulate the CW light with twice larger optical modulation amplitudes (OMA) than that of the LSB segment, due to its longer physical length. Thanks to this ODAC scheme [43–47], the co-integrated MSB and LSB drivers only need to deliver electrical PAM2 signal to the phase shifters for optical PAM4 modulation, relaxing the linearity requirement of the final stage of the CMOS driver. To make the optical path lengths of the upper and lower arms equal, an additional waveguide, with a length of $L_C = 20$ μm, is inserted at the lower optical path. The $n$-$i$-$n$ junction heaters are placed following the MSB segment to control the phase offset.

Detailed circuit schematic and design of the EPIC modulator is shown in Fig. 4. The co-integrated Mach-Zehnder modulator (MZM) driver is composed with two current-mode logic (CML) drivers and one main driver. Thanks to the co-design and integration of the TWE-MZM and driver, the parasitics at the output nodes of the driver are largely reduced [36]. As a result, the shunt-peaking inductors are only applied in CML pre-drivers, enabling wider bandwidth and small area. For impedance matching, the differential output impedance of the main driver is designed to be $2R_0 = 80$ Ω at DC, and the differential TWE signal paths of each modulators are terminated with $2R_0$ polysilicon resistors.

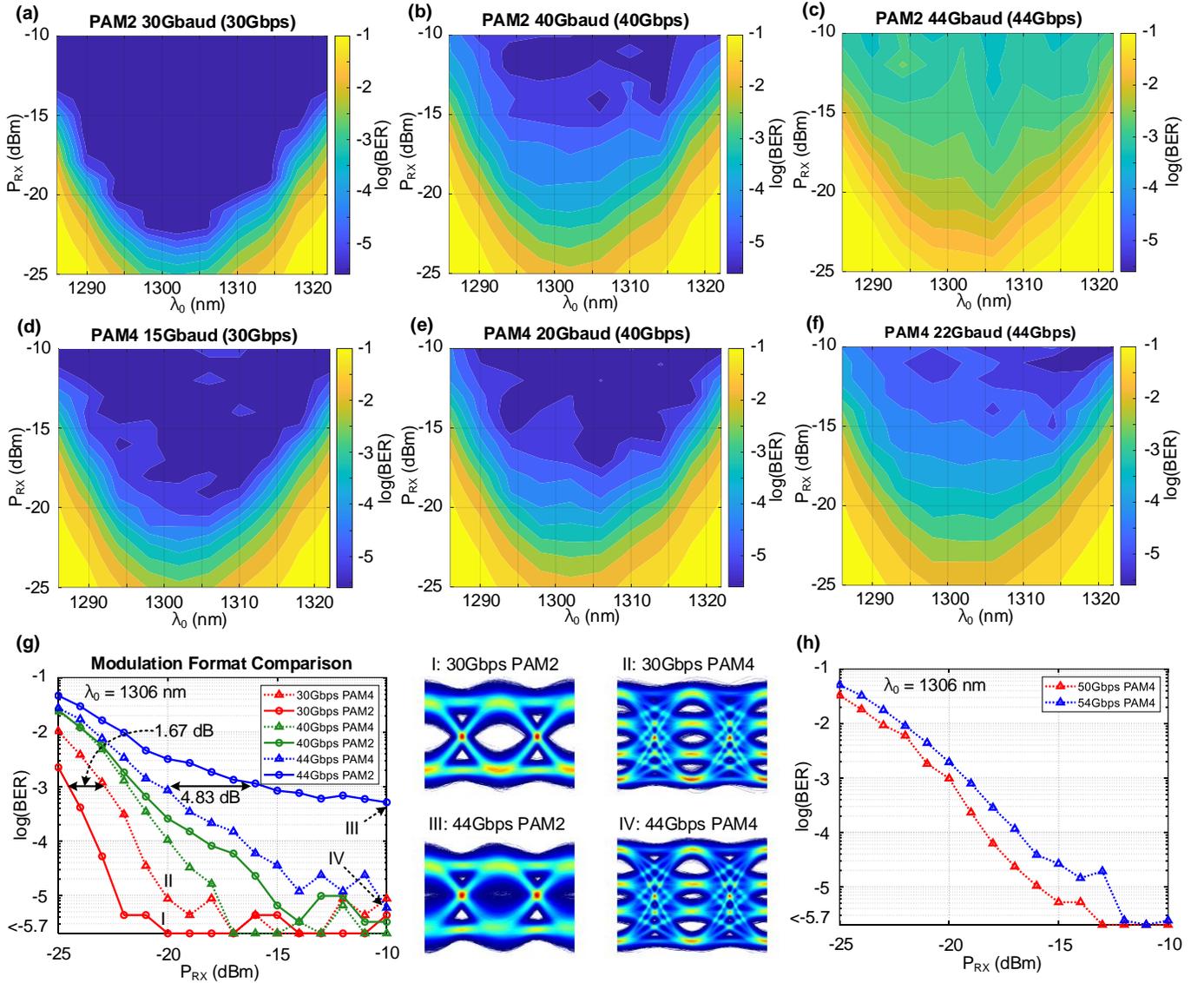

**Fig. 7.** Back-to-back high-speed measurement results of the proposed electronic-photonic integrated circuit (EPIC) modulator. bit error rate (BER) contour with varying received optical power ($P_{RX}$) and the carrier wavelength, when the EPIC modulator is operating with (a)-(c) PAM2 and (d)-(f) PAM4 mode. (g) Comparison of different modulation format and data rates of the proposed EPIC modulator at $\lambda_0 = 1306$ nm. Near 40 Gbps is the singular point where the performance transition between the PAM2 and PAM4 signaling occurs (See the green line). Measured eye diagrams of each data points. (h) BER when the EPIC modulator is operating at the high data rates of 50 and 54 Gbps.

Miller capacitor with a value of 4.8 pF is inserted between the output node of an error amplifier and PMOS current source to compensate the stability of the common-mode feedback loop. Measured DC currents of the first and second CML drivers and main driver are $I_{CML1} = 12$ mA, $I_{CML2} = 24$ mA, and $I_{SS} = 46$ mA, respectively. A microscope image of the presented EPIC modulator is shown in Fig. 5 and comprehensive comparison of the proposed EPIC modulator and prior works is explained in Section III.

The measured normalized output power of the EPIC modulator over the heater voltage, $V_H$, is shown in Fig. 6(a). The result shows that the quadrature point of the EPIC modulator is near $V_H = 0.57$ V. In order to evaluate the modulation capability of the ODAC scheme in our EPIC modulator, we measured the difference between the high-level and low-level output optical power ($P_{OMZM,H}$–$P_{OMZM,L}$) of the modulator as a function of $V_H$ at $\lambda_0 = 1306$ nm, when the input DC voltages of the MSB and LSB segments are independently driven. To measure the high-level optical power ($P_{OMZM,H}$) of the MSB segment, a differential DC voltage of $V_{IND,DC} = V_{INP} - V_{INN} = +1$ V is applied to the input nodes ($V_{INP,MSB}$ and $V_{INN,MSB}$) of the CMOS driver at the MSB segment. Under this condition, $V_H$ is swept from 0 V to 1 V, while the wavelength of the input light is fixed at $\lambda_0 = 1306$ nm. The $P_{OMZM,H}$ is then obtained by observing the optical power ($P_{OMZM}$) emitted from the output GC of the EPIC modulator. To measure low-level optical power ($P_{OMZM,L}$) of the MSB segment, $V_{IND,DC} = -1$ V is applied to $V_{INP,MSB}$, and $V_{INN,MSB}$, and $V_H$ is varied to observe the corresponding $P_{OMZM,L}$. During the measurement of $P_{OMZM,H}$ and $P_{OMZM,L}$ over the $V_H$ $\lambda_0 = 1306$ nm, the input voltages of the CMOS driver at the LSB segment are fixed at $V_{INP,MSB} = V_{INN,MSB} = 0.9$ V. Finally, $P_{OMZM,H}-P_{OMZM,L}$ of the MSB segment as a function of $V_H$ is then calculated by subtracting the measured values of $P_{OMZM,L}$ from $P_{OMZM,H}$. The $P_{OMZM,H}-P_{OMZM,L}$ of the LSB segment is measured through a procedure similar to that used for the MSB segment, but by applying $V_{IND,DC} = +1$ V and -1 V to the input nodes ($V_{INP,LSB}$ and $V_{INN,LSB}$) of the LSB segment's CMOS driver, while maintaining $V_{INP,MSB} = V_{INN,MSB} = 0.9$ V. Fig. 6(b) shows the measurement results of the ($P_{OMZM,H}-P_{OMZM,L}$) of each MSB and LSB segments. Since the MSB phase shifter is two times longer than that of LSB, the change of $P_{OMZM,H}-P_{OMZM,L}$ of the MSB segment with varying $V_H$ is larger than that of LSB. The inset figure shows the difference between the MSB and LSB segments' $P_{OMZM,H}-P_{OMZM,L}$ near the quadrature point, verifying that the ODAC is operating correctly. After the null point at $V_H = 0.86$ V, the measurement results of $P_{OMZM,H}-P_{OMZM,L}$ change to negative value due to a cosine shaped transfer function of the MZM. The effect of different phase shifter lengths is also reflected in the normalized transmissions found for the EPIC modulators, which are shown in Figs. 6(d)-(e). When $V_{IND,DC} = +1$ V and -1 V are applied to the two modulator segments, the transmission spectrum of the MSB segment shows a larger change on the output power than LSB segment. Single-ended EO-S21 were measured by applying a sinusoidal voltage signal, having a peak-to-peak voltage of 200 mV$_{PP}$ and a frequency range between 0.1~25 GHz, to $V_{INP,MSB}$ or $V_{INP,LSB}$, respectively. For these measurements the other three inputs are DC biased with 0.9 V. After amplifying the modulated optical signal from $P_{OMZM}$ through a Praseodymium doped fiber amplifier (PDFA), Fast Fourier Transform is then performed to a time-domain signal detected by a digital communication analyzer to extract the variation of the signal power. Measured single-ended EO-S21 with a -3 dB bandwidth (BW) for the MSB and LSB segments of 18.5 GHz and 20.5 GHz are found, see Fig. 6(c).

Since the transversal angle, $\theta_{//}$, of the MBs are determined by the choice of the carrier wavelengths, a back-to-back (B2B) bit error rate (BER) as a function of the received optical power ($P_{RX}$) of the proposed EPIC modulators have been measured over varying operating wavelengths, $\lambda_0$, as shown in Figs. 7(a)-(f). In order to measure the B2B BER contour over the varying $P_{RX}$ and $\lambda_0$, electrical random data of >230400 bits are fed into our EPIC modulator and encoded on $\lambda_0$. Then, the modulated optical signal from $P_{OMZM}$ is varied and amplified by variable optical attenuator (VOA) and PDFA. Finally, direct detection photodetector (DD-PD) having a BW of 70 GHz and a digital sampling oscilloscope (DSO) recorded the modulated signals. Due to the band limited input and output GCs, the contours show a parabolic shape, where the receiver sensitivities are lowest near the center wavelength of the GCs. Moreover, at lower data rates, lower modulation formats show a better BER performance trend in the contour, while higher order modulation is preferred for higher data rates. To compare the power penalties of the EPIC modulator over different data rates, BERs of PAM2 and PAM4 signals, when the $\lambda_0$ is 1306 nm, are plotted in Fig. 7(g). When the data rate is 30 Gbps, the PAM4 underperforms over PAM2 with a power penalty of 1.67 dB at BER of $1\cdot10^{-3}$. However, the performance of the PAM4 starts to surpass the PAM2, at a data rate near 40 Gbps. Finally, when the data rate is 44 Gbps, the PAM4 outperforms PAM2 with a sensitivity advantage of 4.83 dB. This is mainly caused by the EO-BW limitation of the device, which is determined by parasitic capacitances at the output node of the main driver and RF losses of the TWEs. Fig. 7(h) shows the BERs of our EPIC modulator operating at the high data rates, indicating that the proposed EPIC modulator achieves B2B data rates of up to 54 Gbps.

### C. Multibeam OWC

In order to evaluate the performance of the proposed multibeam OWC architecture, an OWC measurement has been performed by interfacing the proposed MBOPA and EPIC modulator, as shown in Fig. 8. The polarization controlled (PC) CW light with different wavelengths is generated by a tunable laser and coupled into our EPIC modulator. The spacing between the carrier wavelengths, $\lambda_{C,i+1} - \lambda_{C,i}$, is set to be 2 nm and varied from $\lambda_{C,1} = 1304$ nm to $\lambda_{C,10} = 1322$ nm. The proposed EPIC modulates optical carriers, $\lambda_{C,i}$, with the modulation format of PAM2 or PAM4 based on two electrical PAM2 signals, which are fed into the differential inputs, $V_{IND,MSB}$ and $V_{IND,LSB}$, of the MSB and LSB CMOS drivers. Then, our MBOPA creates multibeam by spatially mapping the modulated lights in each $\lambda_{C,i}$ to $\theta_{//,i}$ in free-space. The MB, directing from $\theta_{//,1}$ = +5.57° to $\theta_{//,10}$ = -4.48°, propagates through the one-meter free-space channel and is received by a fiber collimator (FC) with a clear aperture of 42.5 mm. Lastly, the received optical power, $P_{RX}$, is varied and amplified by the VOA and PDFA, and detected by the 70 GHz-BW DD-PD and DSO.

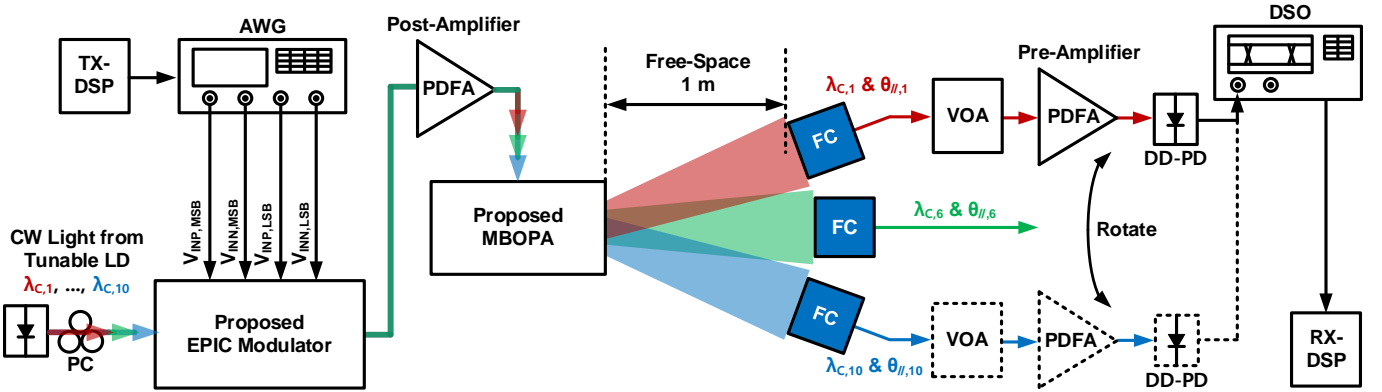

**Fig. 8.** Measurement setup to characterize OWC performance using ten wavelengths.

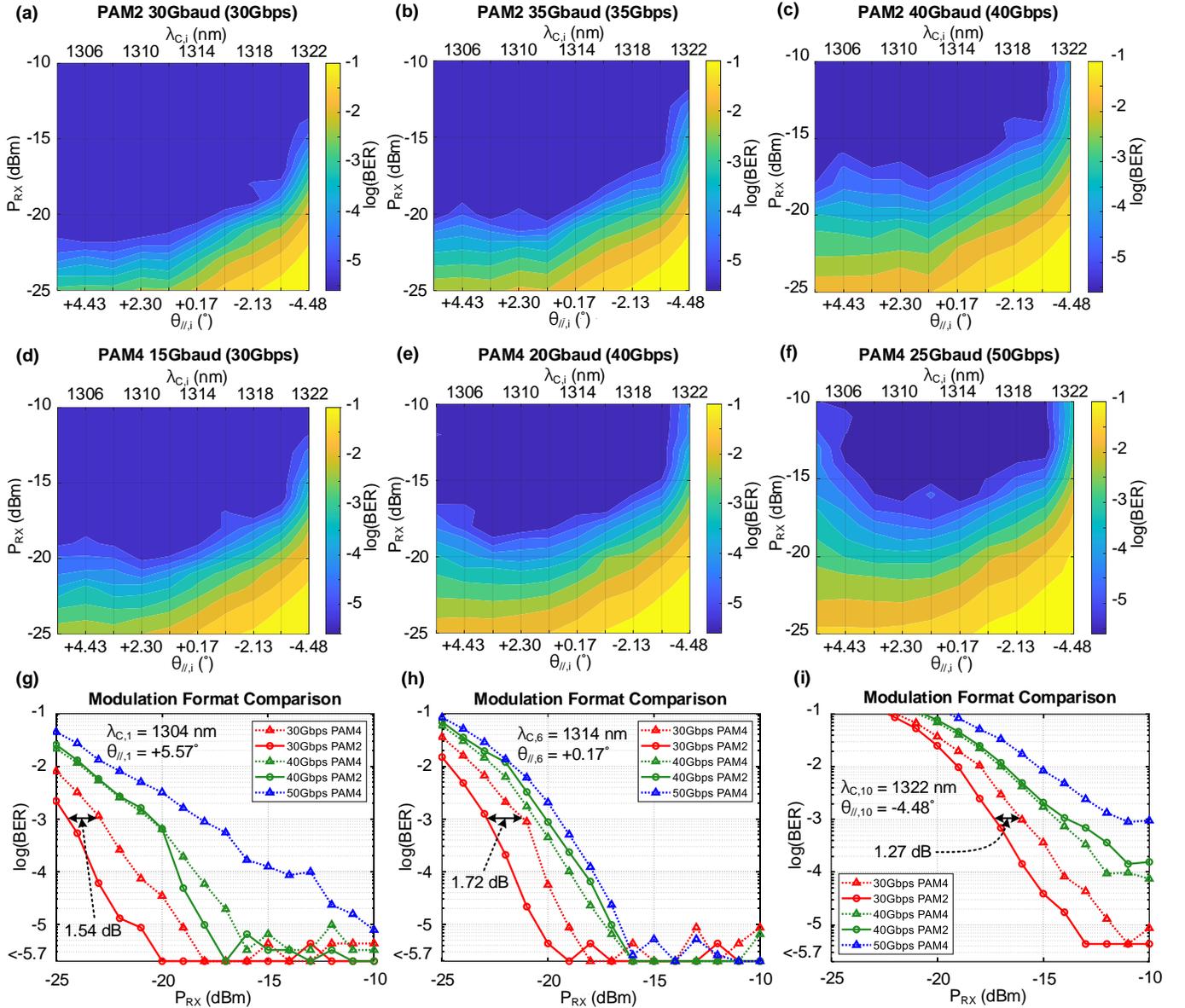

**Fig. 9.** Free-space BER contour with varying received optical power ($P_{RX}$) over the different transversal angles of MB, $\theta_{//,i}$, when the proposed multi-user OWC system is operating with (a)-(c) PAM2 and (d)-(f) PAM4 mode. Comparison of different modulation format and free-space data rates of the proposed OWC system of the chosen MB, (g) $\theta_{//,1}$ = +5.57°, (h) $\theta_{//,6}$ = +0.17° and (i) $\theta_{//,10}$ = -4.48°.

Figs. 9(a)-(f) show the one-meter OWC measurement results using ten multibeams plotted as BER contours as a function of the $P_{RX}$ and $\theta_{//,i}$. The right side ($\lambda_{C,10}$ = 1322 nm, $\theta_{//,10}$ = -4.48°) of the contours show rapid degradation on the BER due to the optical BW limitations EPIC modulator and MBOPA. Since the measured -3 dB BW through the transmission spectrum of the EPIC ranges from 1288 nm to 1319 nm, see Figs. 6(d)-(e), and -2.3 dB attenuation of the far-

field intensity of the MBOPA occurs near 1304 nm and 1325 nm, see Fig. 2(c), the modulated signal on $\lambda_{C,10}$ = 1322 nm suffers from both of these effects. On the other hand, left side ($\lambda_{C,1}$ = 1304 nm, $\theta_{//,1}$ = +5.57°) of the contours is mainly affected by the MBOPA's far-field envelope, since the center wavelength of the EPIC is near 1306 nm. The PAM2 signals show great resilience on the effect of the far-field envelope in the high baud rates, while the BER of the PAM4 signal on $\lambda_{C,1}$ = 1304 nm worsens as the baud rates becomes higher. To compare the modulation format and data rates of the proposed multi-user OWC architecture, the measured BER over $P_{RX}$ at $\theta_{//,1}$ = +5.57°, $\theta_{//,6}$ = +0.17°, $\theta_{//,10}$ = -4.48° is plotted in Figs. 9(g)-(i). In the lower data rates, PAM2 performs better than PAM4 with a sensitivity advantage up to 1.72 dB. However, at a data rate of 40 Gbps, the performance of the PAM2 and PAM4 signal becomes similar, and only PAM4 can support a data rate of 50 Gbps in the OWC link.

To evaluate the power penalty when the number of users in the proposed OWC system is increased, we configured the experimental setup as shown in Fig. 10(a). Due to the limited physical size of the optical table, only one probe station for the new EPIC modulator is used. Consequently, the single modulator by multiplexing two CW lights as its input is employed to generate the WDM signal, instead of using multiple EO-modulators as shown in Fig. 1. Two CW lights from the laser A and B, each having a wavelength of $\lambda_{C,A} = 1305$ nm and $\lambda_{C,B} = 1318$ nm, are first multiplexed using a 50:50 coupler and then amplified by a silicon optical amplifier (SOA). The spacing between $\lambda_{C,A}$ and $\lambda_{C,B}$ was determined by the minimum distance between the alignment stages of the two FCs (RX-A and RX-B) in Fig. 10(a). The proposed EPIC modulator encodes PAM2 or PAM4 signals from the AWG onto CW carrier wavelengths, which are $\lambda_{C,A}$ and $\lambda_{C,B}$. Finally, the BER of the modulated light on $\lambda_{C,B}$ is measured by the FC positioned at the RX-B, $\theta_{//,B} = -2.13°$, while the signal on $\lambda_{C,A}$ is aligned to RX-A, $\theta_{//,A} = +5.06°$.

Figs. 10(b)-(c) show the PAM2 and PAM4 signal's BER, and Fig. 10(d) illustrates the BER at the highest measurable data rate. The blue curves in Figs. 10(b)-(d) represent the measured BER at RX-B when the $\lambda_{C,B}$ laser is turned on, while the red curves show the BER when both $\lambda_{C,A}$ and $\lambda_{C,B}$ lasers are on. In Fig. 10(a), both the SOA and PDFA exhibit gain

saturation effect when the laser A and B are on [49, 50]. However, in a practical multiuser OWC deployment, the SOA is not used, see Fig. 1. Hence, an estimated effect of the SOA's gain saturation on the power penalty is de-embedded (See details in Section II-D). The power penalty due to the gain saturation effect in the post-amplifier is 3.99 dB. Hence, the transmit power should be adjusted to support simultaneous communication with two users.

### D. De-embedding

To de-embed the effect of the SOA on the power penalty, the OWC link in Fig. 10(a) is simplified as a block diagram in Fig. 11. The block diagram in Fig. 11 consists of three cascaded optical amplifiers and the optical power losses from the modulator, $L_{MZM} = 18$ dB, and from free-space channel to FC at RX, $L_{CH} = 20.7$ dB.

In a system with cascaded optical amplifiers, noise is dominated by amplified spontaneous emission (ASE) [51]. Thus, optical signal-to-noise ratio (OSNR) at the output of the Stage-i amplifier ($OSNR_i$) can be described as:

$$OSNR_i = \frac{1}{\frac{1}{OSNR_{i-1}} + \frac{NF_i hf B_r}{P_{IN_i}}} \qquad (1)$$

where $OSNR_{i-1}$ is the OSNR at the output of the preceding stage's amplifier, $h$ is Planck constant, $NF_i$ is the noise figure, and $P_{IN_i}$ is the input power of the current stage's amplifier. The frequency of the light and resolution bandwidth are set to be $f$ = 227.5 THz and $B_r$ = 17.3 GHz, respectively. Hence, the final OSNR seen following the DD-PD at the RX can be expressed as follows:

$$OSNR_3 = \frac{1}{\frac{1}{OSNR_1} + \frac{NF_2 h f B_r}{\frac{P_{O,SOA}}{L_{MZM}}} + \frac{NF_3 h f B_r}{\frac{P_{O,SOA} G_2}{L_{MZM} L_{CH}}}} \quad (2)$$

where $P_{O,SOA}$ is signal power at the output of the SOA in Stage-1 and $G_2$ is the gain of the PDFA in Stage-2. According to Eq. (2), the $OSNR_3$ can be calculated by measuring $P_{O,SOA}$ and $OSNR_1$ at $\lambda_{C,B}$ = 1318 nm using optical spectrum analyzer.

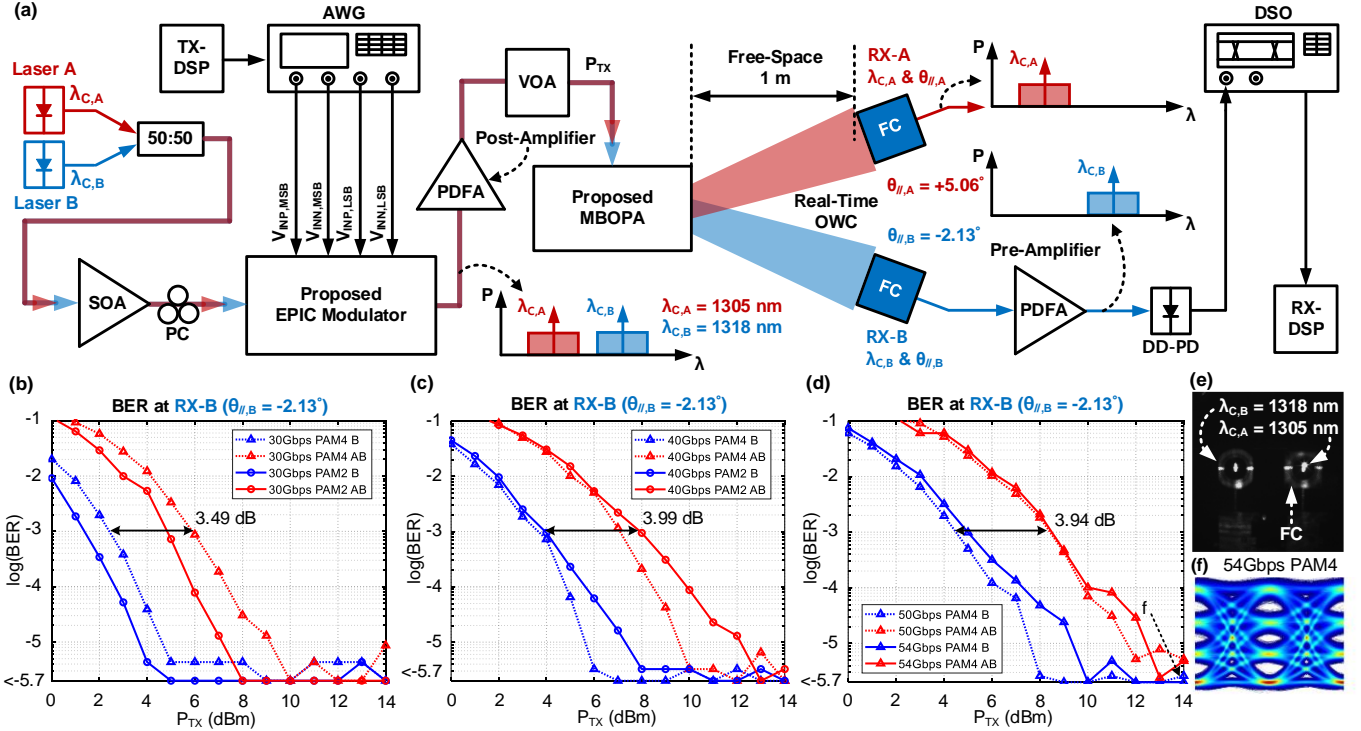

**Fig. 10.** (a) Measurement setup to evaluate the effect of wavelength division multiplexing (WDM) signal in the multi-user optical wireless communication (OWC) system. The spacing between the two wavelengths, $\lambda_{C,A}$ and $\lambda_{C,B}$, are limited by the physical size of the fiber collimator (FC) stages and distance between the transmitter (TX) and receiver (RX). Measured bit error rate (BER) at the RX-B ($\theta_{//,B} = -2.13°$), over the varying input power to the MBOPA (TX), when the data rate is (b) 30 Gbps, (c) 40 Gbps, and (d) 50/54 Gbps. Blue and red curves are the BER, when the laser A is turn on and both laser A and B are turned on, respectively. (e) Photograph of the two FCs at the RX taken by infrared camera. (f) Eye diagram of the 54 Gbps PAM4 signal at $P_{TX} = 17$ dBm, when the two lasers are turned on.

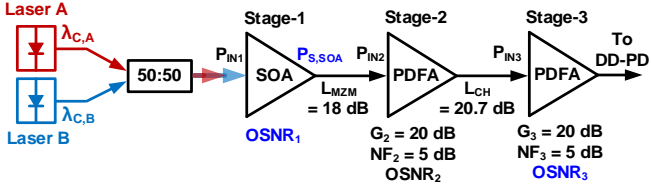

**Fig. 11.** Simplified block diagram of the OWC system in Fig. 10(a), consisting of three-stage optical amplifiers.

TABLE I
STAGE-1 MEASUREMENT RESULTS AND ITS CONTRIBUTION
TO THE SYSTEM POWER PENALTY

| Parameters at $\lambda_{C,B}$ | Laser B On | Laser A & B On |
|---|---|---|
| Signal Power ($P_{S,SOA}$) | 1.93 dBm | 0.17 dBm |
| Noise Power ($P_{N,SOA}$) | -49.15 dBm | -50.67 dBm |
| $OSNR_1$ | 51.08 dB | 50.84 dB |
| $OSNR_3$ | 31.35 dB | 29.60 dB |
| System Power Penalty | 1.75 dB | |

Table I shows the summarized measurement results of $P_{O,SOA}$, $OSNR_1$, calculated $OSNR_3$ at $\lambda_{C,B}$ = 1318 nm, when only Laser B is on and on both Laser A and B are on, and the power penalty that needs to be de-embedded for the practical deployment of the OWC system. The power penalty caused by injecting multiplexed CW lights into the SOA is evaluated by subtracting the two $OSNR_3$ values, which are obtained by turning on and off Laser A, and the de-embedded results are shown in Figs. 10(b)-(d).

## III. DISCUSSION

Table II shows performance summary of the proposed EPIC modulator and comparison with previously published MZMs and drivers, which are monolithically integrated. Compared to other designs, our proposed EPIC achieves high data rates, low power consumption, and wide output swing at the same time. Reference [47] achieves a higher data rate thanks to a common source topology in its main driver. However, the differential output swing is only 2.16 $V_{PP}$ limited by the voltage stress limit of the transistors and the pull-down only structure. On the contrary, thanks to our cascode push-pull topology, our proposed EPIC modulator achieves the driver output swing of 3.2 $V_{PP}$, which leads to higher OMA than [47].

The performance of the presented EP architecture for multi-user OWC is shown in Table III., which are compared with performances of other state-of-the-art works. Since [32–33, 35] all use off-the-shelf discrete MZMs and RF-amplifiers, the estimated power consumption of the commercial components are used for the comparison, as described in [36].

Our proposed EP interface for OWC is the world's first OPA TX that generates MBs to support multi-user communication. Through the proposed high-performance EPIC modulator, MBOPA with AWG delay lines, and WDM technique, the system can support parallel communication with two users over the free-space communication distance of 1 m, resulting a total communication capacity of 108 Gbps, where the number of users and communication distance are limited by the pitch of the antenna, size of the optical table and alignment stages for the FCs. Due to the limited number of the PDFAs at the RX system, the signals were separately detected, although the FCs for RX-A and RX-B simultaneously receive the signals modulated in the multibeams. The area of our proposed OWC system, which is capable of creating MB having a beam divergence angle of 0.79°, is only 186.8 mm², including one MBOPA and two EPIC modulators, which is significantly smaller than the RF MBPAAs. For comparison, a previously published Ka-band lens-based PMBA, supporting seven simultaneous beams with a beam width of 6.6°, occupies 10241 mm² [5].

Thanks to the WDM and AWG-based MB generation techniques, the proposed MB system offers high scalability in terms of the number of beams. While the size and complexity of the conventional RF and mm-wave PMBAs and MBPAAs grow significantly with the number of required beams [3], our proposed MBOPA maintains its size. However, the upper limit of the MBOPA system scalability is governed by two factors: the angular spacing between adjacent MBs and the gain that the optical post-amplifier at the transmitter can apply across multiple carrier wavelengths. Since the average value of the measured beam steering angle and free spectral range of the MBOPA are 12.18° and 21.4 nm, respectively, the EP system can potentially support OWC with four users, when the communication distance is 10 m and the nearest neighboring RXs are spaced with 50 cm, which together achieve calculated communication capacity of 216 Gbps.

Finally, we summarize the characteristics of different modulation formats of the proposed EP system for OWC as follows. Lower order modulation is preferred when the OWC link requires lower data rate and large number of users. As shown in Fig. 9, PAM2 signal outperforms PAM4 signal in a data rate less than 40 Gbps, while the opposite case happens at the higher data rates, due to the limitation of the EO BW of the proposed EPIC modulator, see Fig. 6(c). Moreover, the PAM2 signal has less vulnerability on the effects of the far-field intensity attenuation of the MBOPA at wider angles. Near the pair of carrier wavelength and angle where the far-field intensity of the MB reduces down to -2.3 dB (i.e. $\lambda_{C,1}$ = 1304 nm, $\theta_{//,1}$ = +5.57° in Fig. 9), PAM2 signal shows better BER profile than that of PAM4 signal, see Figs.9 (b), (c), (e), and (f). Hence, when the OWC system aims to communicate with large number of users in a wide FoV, PAM2 signal is preferred. On the contrary, if the communication system requires a high data rate up to 54 Gbps, PAM4 modulation should be chosen at the cost of the number of simultaneous users.

TABLE II
COMPARISON TO STATE-OF-THE-ART MONOLITHIC MZM AND DRIVER

| Publication | **This work** | Optica 2016 [47] | CLEO 2021 [48] | JLT 2024 [36] |
|---|---|---|---|---|
| Technology | **45-nm Monolithic** | 90-nm Monolithic | 90-nm Monolithic | 45-nm Monolithic |
| MZM Topology | **2-bit TWE-ODAC** | 2-bit TWE-ODAC | Multi Segment | TWE |
| Driver Topology | **Cascode Push-Pull** | Cascode CS | Cascode CS | Cascode Push-Pull |
| Data Rate (Gbps) | **54** | 56 | 30 | 15 |
| Supply Voltage (V) | **1.5/3.0** | 1.2/1.5 | 3.3 | 1.5/3.3 |
| Power (mW) | **384** | 270 | 380 | 210 |
| Output Swing ($V_{PP}$) | **3.20[*]** | 2.16[*] | N/A | 3.20[*] |
| Active Area (mm²) | **3.4** | 3.75 | 4.03 | 1.43 |
| FOM1 (mW/Gbps) | **7.11** | 4.8 | 12.7 | 14 |
| FOM2 (mW/Gbps/$V_{PP}$) | **2.22** | 2.23 | N/A | 64 |

[*]: Simulation result.

[†]: FOM2 = Power/Data Rate/Output Swing.

TABLE III
COMPARISON TO STATE-OF-THE-ART OPAS FOR OWC

| Publication | **This work** | JSTQE 2019 [32] | PTL 2020 [33] | JLT 2023 [35] | JLT 2024 [36] |
|---|---|---|---|---|---|
| Material | **Silica** | Silicon | Silicon | Silicon | Silicon |
| Wavelength | **O-band** | C-band | C-band | C-band | O-band |
| Multi-user OWC | **Yes** | No | No | No | No |
| The Number of the Users | **2[††]** | 1 | 1 | 1 | 1 |
| OWC Data Rate (Gbps) | **108[††]** | 10 | 32 | 50 | 2 |
| MZM & Driver Power (mW) | **386** | 3120[†] | 6360[†] | 6360[†] | 210 |
| The Number of the Array | **101** | 512 | 64 | 128 | 64 |
| Side-lobe Suppression Ratio (dB) | **12.6** | 12 | 12 | 10.63 | 7.81 |
| Beam Divergence $\theta_{//}$, $\theta_{\perp}$ (°) | **0.79, 0.18** | 0.04, 0.04 | 0.7, 0.9 | 0.02, 0.03 | 0.77, 4.23 |
| Steering Angle $\theta_{//}$, $\theta_{\perp}$ (°) | **12.18** | 54, 15.0 | 46, 10.2 | 100, 2.1 | 28.6, 6.1 |
| Size of the OPA (mm²) | **180[**]** | 10[*] | 16.7[**] | 12[*] | 6.4[**] |

[*]: Size of the antennas.

[**]: Size of the OPA chip.

[†]: Estimated power of commercial MZM and RF-amplifiers. Details are presented in [36].

[††]: Separately measured due to the limited number of PDFAs.

## IV. CONCLUSION

An electronic photonic interface using a novel EPIC modulator, MBOPA, and WDM technique, which can create multi-beam and communicate with multi-users is presented. Our high-speed EPIC modulator, which is fabricated by GlobalFoundries' 45-nm monolithic silicon photonics technology (45SPLCO), achieves measured data rates of up to 54 Gbps and only consumes 386 mW. In addition, the 2-bit ODAC structure in the EPIC modulator not only provides reconfigurability on the modulator formats but also relaxes the linearity requirement of the co-integrated CMOS driver. The proposed silica-based MBOPA with AWG delay lines simultaneously directs WDM data signals to different angular directions, enabling a massive MIMO communication. According to the measurement results, our architecture can communicate with two users, providing the total data rate of 108 Gbps, when the wireless communication distance is 1 m, which is limited by the pitch of the MBOPA and measurement setup. We also analyze the performance of the different modulation formats that the EPIC modulator can provide. PAM2 signal is preferred for the OWC link requiring relatively low-speed and large number of users. The proposed architecture opens up the way for the future wireless communication beyond 6G that can overcome the fundamental issues of the conventional RF and mm-wave systems.

## FUNDING

This work was supported in part by ETH Zürich Grant, Swiss State Secretariat for Education, Research, and Innovation (SERI) through SwissChips Initiative.

## ACKNOWLEDGMENT

The authors would like to thank GlobalFoundries for providing silicon fabrication through the 45SPCLO university program. We also thank members at Integrated Devices, Electronics, and Systems (IDEAS) group at ETH Zürich for their valuable discussions and technical supports. We further thank WayOptics Co. for their technical support with the silica MBOPA chip.

**Youngin Kim** (Student Member, IEEE) received B.S. and M.S. degree in electrical engineering from Korea University, Seoul, and Korea Advanced Institute of Science and Technology (KAIST), Daejeon, South Korea, in 2017 and 2019, respectively. He is currently working toward the Ph.D. degree in the Department of Information Technology and Electrical Engineering (D-ITET), Swiss Federal Institute of Technology Zürich (ETH Zürich), Zürich, Switzerland. His research interests include integrated circuits and silicon photonics.

**Laurenz Kulmer** received the B.Sc. and M.Sc. degrees in electrical engineering and information technology in 2019 and 2021, respectively, from ETH Zurich, Zurich, Switzerland, where he is currently working toward the Ph.D. degree with the Institute of Electromagnetic Fields, Zurich, Switzerland. His research interests include optical communication, subTHz communication systems, and subTHz sensing systems.

**Jae-Yong Kim** received his B.S. degree in EE from Yonsei University, South Korea, in 2018. He is currently pursuing a Ph.D. in EE at KAIST, with graduation expected in 2025. From 2018 to 2020, he worked as an engineer at Samsung Display. His research interests include silicon photonics, nanophotonics, and optical communication.

**Hamza Kurt** received the B.S. degree in Electrical and Electronics Engineering from Middle East Technical University, Ankara, Turkey, in 2000; the M.S. degree from the University of Southern California, Los Angeles, USA, in 2002; and the Ph.D. degree from the Georgia Institute of Technology, Atlanta, USA, in 2006. He conducted postdoctoral research at the Institut d'Optique Graduate School in Palaiseau, France, before joining TOBB University of Economics and Technology, Ankara, Turkey, where he held academic roles including Department Head. He is currently a tenured Associate Professor at the Korea Advanced Institute of Science and Technology (KAIST), Daejeon, Republic of Korea, where he leads the Metaphotonics Research Group. His research interests include nanophotonics, metasurfaces, silicon photonics, inverse design, and photonic computing. He is a full member of the Turkish Academy of Sciences and has received several distinctions, including the Turkish Academy of Sciences Distinguished Young Scientist Award (2010) and the KAIST Outstanding Teaching Award (2024).

**Juerg Leuthold** (Fellow, IEEE) received the Ph.D. degree in physics from ETH, Zürich, Switzerland, in 1998, for work in the field of integrated optics and all-optical communications. Since 2013, he has been the Head of the Institute of Electromagnetic Fields, ETH Zurich, Zurich, Switzerland. From 2004 to 2013, he was with the Karlsruhe Institute of Technology, Karlsruhe, Germany, where he was the Head of the Institute of Photonics and Quantum Electronics and the Helmholtz Institute of Microtechnology. From 1999 to 2004, he was with Bell Labs, Lucent Technologies, Holmdel, NJ, USA, where he performed device and system research with III/V semiconductor and silicon optical bench materials for applications in high-speed telecommunications. His research interests include photonics, plasmonics, and microwave with an emphasis on applications in communications and sensing. Dr. Leuthold is a Fellow of the Optical Society of America and a Member of the Heidelberg Academy of Science. He was with the community as a Member of the Helmholtz Association Think Tank, Board of Directors of OSA, and in many technical program committees or as the General Chair of meetings.

**Hua Wang** (Fellow, IEEE) received the M.S. and Ph.D. degrees in electrical engineering from the California Institute of Technology, Pasadena, CA, USA, in 2007 and 2009, respectively. He was a Tenure-Track Assistant Professor and then a Tenured Associate Professor at the School of Electrical and Computer Engineering (ECE), Georgia Institute of Technology, Atlanta, GA, USA, from 2012 to 2021, where he held the Demetrius T. Paris Professorship with the School of ECE. He was the Founding Director of the Georgia Tech Center of Circuits and Systems (CCS), Atlanta, and the Director of the Georgia Tech Electronics and Micro-System (GEMS) Laboratory, Atlanta. Prior to that, he was with Intel Corporation, Santa Clara, CA, USA, and Skyworks Solutions, Irvine, CA, USA, from 2010 to 2011. In 2021, he joined the Faculty of the Swiss Federal Institute of Technology Zürich (ETH Zürich), Zürich, Switzerland, where he is a Full Professor and the Chair of electronics with the Department of Information Technology and Electrical Engineering (D-ITET). He is also the Director of the ETH Integrated Devices, Electronics, and Systems (IDEAS) Group. He has authored or coauthored over 200 peer-reviewed journals and conference papers. He is interested in innovating analog, mixed-signal, RF, and millimeter-wave (mm-Wave) integrated circuits and hybrid systems for wireless communication, sensing, and bioelectronics applications.